\documentclass[aip,amsmath,amssymb,reprint,groupedaddress]{revtex4-1}

\usepackage{graphicx}
\usepackage{dcolumn}
\usepackage{bm}
\usepackage[utf8]{inputenc}
\usepackage[T1]{fontenc}
\usepackage{mathptmx}
\usepackage{etoolbox}
\usepackage{xcolor}
\DeclareSymbolFont{calletters}{OMS}{cmsy}{m}{n}
\DeclareSymbolFontAlphabet{\mathcal}{calletters}



\begin{document}

\preprint{AIP/123-QED}

\title[Egg-speriments: Stretch, crack and spin]{Egg-speriments: Stretch, crack, and spin}

\author{Yann Bertho}
\email{yann.bertho@universite-paris-saclay.fr}

\author{Baptiste Darbois Texier}
\author{Ludovic Pauchard}
\affiliation{Universit\'e Paris-Saclay, CNRS, Laboratoire FAST, 91405 Orsay, France}

\begin{abstract}
Eggs are key ingredients in our kitchens because of their nutritional values and functional properties such as foaming, emulsifying and gelling, offering a wide variety of culinary achievements. They also constitute ideal objects to illustrate a myriad of scientific concepts. In this article, we focus on several experiments (egg-speriments) that involve the singular properties of the liquids contained inside the eggshell, especially the egg white. We first characterize the rheology of an egg white in a rotational rheometer for constant and oscillatory shear stresses revealing its shear-thinning behavior and visco-elastic properties. Then, we measure the tendency of the fluid to generate very long filaments when stretched that we relate to the shear modulus of the material. Second, we explore the anisotropic crack pattern that forms on a thin film of egg white after it is spread on a surface and let dried. The anisotropy results from the long protein chains present in the egg white which are straightened during film deposition. Finally, we consider the ``spin test'' that permits to distinguish between raw and hard-boiled eggs. To do so, we measure the residual rotation of a spinning raw egg after a short stop which reflects the continuation of the internal flow. These observations are interpreted in terms of viscous damping of the internal flow consistently with the measurements deduced from rheology.
\end{abstract}

\maketitle

\section{Introduction}

A chicken egg, which constitutes an essential ingredient of our kitchens, is made of different parts, including the shell with an average thickness of 0.36~mm, \cite{sun2012global} the egg white, and the egg yolk that constitutes approximately one-third of the egg volume \cite{atilgan2008rheological}. The two liquids contained in the eggshell show a great versatility of behaviors: the egg white forms very long filaments when poured, turns to foam when beaten (eggs snow), and gelifies when cooked. The yolk leads to an emulsion when mixed with oil that is at the base of mayonnaise sauce. This rich phenomenology of behaviors makes eggs particularly attractive for culinary preparations \cite{haumont2020chimiste}.

From a scientific point of view, the raw egg white solution is a colloidal suspension formed primarily by $10\%$ proteins (148 proteins of different types of which ovalbumin is the most abundant) dissolved into $90\%$ water \cite{strixner2011egg}. These long proteins form a three-dimensional entangled network. The quality of this network depends mainly on the physico-chemical conditions of the medium (specifically pH, ionic strength, and type of salts) \cite{gossett1983selected}.
The properties of these complex fluids are a major concern in food industries, which seek for the correct processing and equipment design to deal with egg products \cite{polachini2017rheology} and continue to fascinate the scientists who seek to refine their understanding of these behaviors.
For example, recently, Begam \textit{et al.} used coherent x-ray scattering to probe the kinetics of heat-induced gelation and the microscopic dynamics of a hen egg white gel \cite{begam2021} and revealed that the kinetics of the structural growth is a reaction-limited aggregation process with a gel fractal dimension close to 2, leading to a gel network of average mesh size of 400~nm.
\begin{figure}[b]
\includegraphics[width=\hsize]{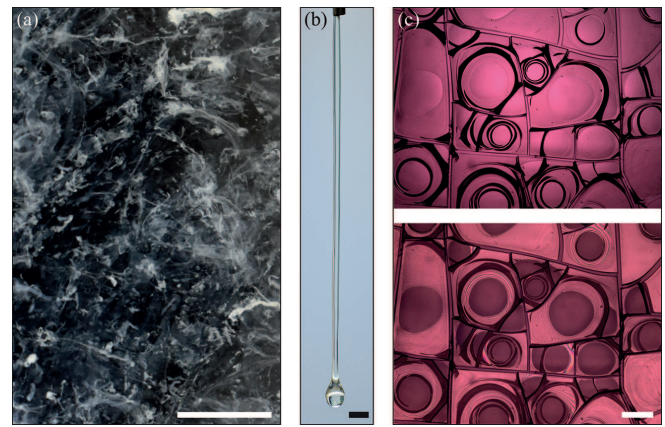}
\caption{(a)~Filamentous structures observed when a drop of egg white is diluted in still water. The scale bar represents 10~mm. (b)~Static shape of an egg white pendant drop. The filament aspect ratio (length over radius) is larger than 50. The scale bar represents 2~mm. (c)~Crack patterns induced by the drying of an egg white film deposited on a glass slide. The scale bar represents 200~$\mu$m. Top: image observed in transmitted light. Bottom: image observed in reflected light microscopy.}
\label{Fig01}
\end{figure}

A simple egg in a kitchen allows to illustrate many scientific concepts. For example, the rise of a hard-boiled egg from horizontal to vertical when spun sufficiently rapidly on a table allows to discuss concepts of solids mechanics \cite{moffatt2002spinning,cross2018does}. Also, an egg spinning through a pool of milk on the kitchen counter and inducing milk rise and droplet ejection constitutes a nice example of fluid mechanics and capillary phenomenon \cite{langley2015eggs}. Among the possible experiments that can be realized with an egg, some of them involve the complex fluids contained inside the eggshell. This is the case when we separate the yolk from the egg white : because of its ``filamentous aspect'' [Fig.~\ref{Fig01}(a)], the white remains attached to any eggshell asperity, is submitted to self-siphoning when poured into a bowl, and is able to generate extremely long filaments when stretched by gravity [Fig.~\ref{Fig01}(b)]. A second example is when a film of egg white is spread on a surface and dries, it induces very specific crack patterns that are characteristic of a brittle material. Figure~\ref{Fig01}(c) displays the patterns induced by the drying of an egg white film deposited on a glass slide. The top image, obtained in transmitted light, exhibits cracks that divide the film into polygonal cells possibly enclosing circular or spiral cracks. The bottom image, obtained in reflected light microscopy, displays optical interference fringes that reveal regions of de-adhesion at the film/substrate interface (light zones reflecting air gap) and regions where the film adheres to the substrate (dark circular zones). Such morphologies are usually encountered during the drying of films made of organic or inorganic components as well as biological solutions \cite{Lazarus2011,YaPu2018,Pauchard2020}. In particular, the topology of the crack network is the signature of the microstructure. Another situation where we experience the specific flowing behavior of the liquids contained in the eggshell is the spin test that permits to distinguish a raw egg from a hard-boiled egg. This test consists in placing the egg on a hard surface, like the counter, and spinning it like a top. As it spins, grab it with your fingers ever-so-briefly and immediately let go. If the egg keeps spinning, it is raw: indeed, the fluid inside the egg continues to rotate and is able to put the egg back in motion. On the other hand, if the egg stops dead, it is boiled.

In this article, we focus on the physics of these situations where the flow of the complex liquids contained inside the eggshell plays a major role. First, we quantify the flowing behavior of an egg white in a rheometer. Then, we measure the elongation of a hanging filament of egg white related to its visco-elastic properties. Thereafter, we realize velocity-controlled film deposition of egg white, and we characterize the pattern formed after drying as a function of the film thickness. Finally, we reproduce a controlled spin test at the laboratory to analyze the residual rotation of a raw egg that can be explained with a simple viscous damping model, in agreement with the previous rheological measurements.

\section{Rheology}

Rheological measurements of egg white are performed using a rotational rheometer (Anton Paar MCR 501) equipped with a cone-plane geometry (angle $1\,^{\circ}$ and diameter 60~mm). As we do in our kitchen, we separate the white from the yolk of a fresh egg, and we measure the rheology of the whole white. A thermostatic bath is used to maintain the sample temperature at $T=20\,^{\circ}$C.

\begin{figure}[t]
\centering
\includegraphics[width=\hsize]{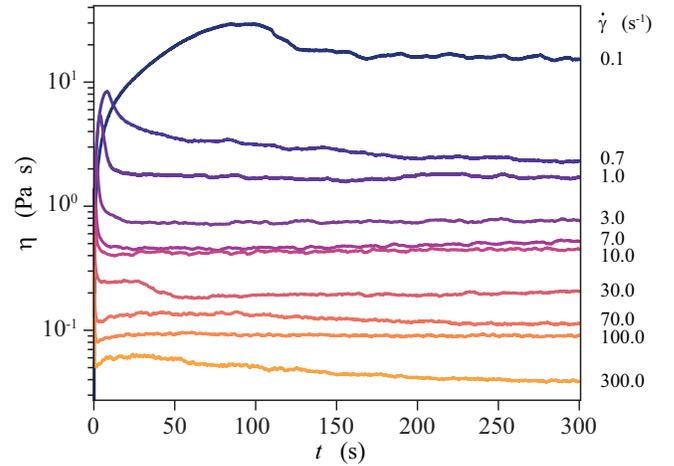}
\caption{Egg white viscosity $\eta$ as a function of time $t$ for different imposed shear rates $\dot{\gamma}$ in the range 0.1~s$^{-1}$ -- 300~s$^{-1}$.}
\label{Fig02}
\end{figure}

Figure~\ref{Fig02} presents the dynamic viscosity $\eta$ of the egg white as a function of time for different imposed shear rates $\dot{\gamma}$ between 0.1 and 300~s$^{-1}$.
After a transient regime, the fluid viscosity is observed to reach a steady value $\bar{\eta}$ that remains roughly constant during the duration of our experiments (5~min). Thus, at this time scale, the fluid does not exhibit clear thixotropic behaviors. However, these results reveal that the white egg is a non-Newtonian fluid as the viscosity plateau strongly depends on the value of the imposed shear rate.

\begin{figure}[t]
\centering
\includegraphics[width=\hsize]{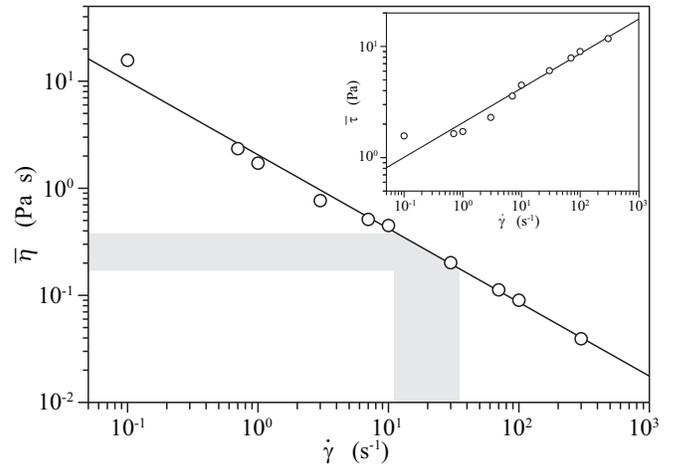}
\caption{Steady viscosity $\bar{\eta}$ of egg white as a function of the shear rate $\dot{\gamma}$. Inset: steady shear stress $\bar{\tau}$ as a function of the shear rate $\dot{\gamma}$. Solid lines correspond to the best fits of the data with the rheological law Eq.~(\ref{eq:rheology}) found for $n=0.31$ and $k=2.1$ Pa s$^{-n}$. The gray region highlights the range of $\dot{\gamma}$ explored in the spin test experiment (see Sec.~\ref{sec:spin}).}
\label{Fig03}
\end{figure}

Figure~\ref{Fig03} shows the steady viscosity $\bar{\eta}$ as a function of the shear rate $\dot{\gamma}$, and the inset in Fig.~\ref{Fig03} presents the corresponding steady shear stress $\bar{\tau}$ as a function of $\dot{\gamma}$. We observe that the fluid viscosity strongly decreases with shear rate, while the shear stress increases with $\dot{\gamma}$. At $\dot{\gamma}=1$~s$^{-1}$, the egg white is about two thousand times more viscous than water, and the viscosity ratio drops to about four hundred times at $\dot{\gamma}=10$~s$^{-1}$. This shear-thinning behavior can be rationalized with a power-law rheology that expresses
\begin{equation}
\bar{\tau} = k\ \dot{\gamma}^{n},
\label{eq:rheology}
\end{equation}
or equivalently, with a viscosity $\bar{\eta} = k\ \dot{\gamma}^{n-1}$. The best fits of our data with this model are found for $n=0.31 \pm 0.01$ and $k=2.1 \pm 0.1$~Pa~s$^{-n}$ as shown in Fig.~\ref{Fig03} with solid lines. These experiments have been realized with different fresh eggs, and a good reproducibility has been observed whatever the egg origin (within 15\% of dispersion). The shear-thinning behavior and the low yield stress for the egg white have also been reported previously by Polachini \emph{et al.} who performed measurements in a concentric cylinder geometry \cite{polachini2017rheology}. In this study, we do not follow a specific procedure to determine precisely the yield stress of the egg white. However, our data reveal that this yield stress is lower than $0.8$ Pa, in good agreement with the values measured by Razi \emph{et al.} \cite{Razi2020}. In the following, we focus on situations where the egg white is submitted to stresses much larger than this value of the yield stress, and we neglect its impact of the considered phenomena.

\begin{figure}[t]
\centering
\includegraphics[width=\hsize]{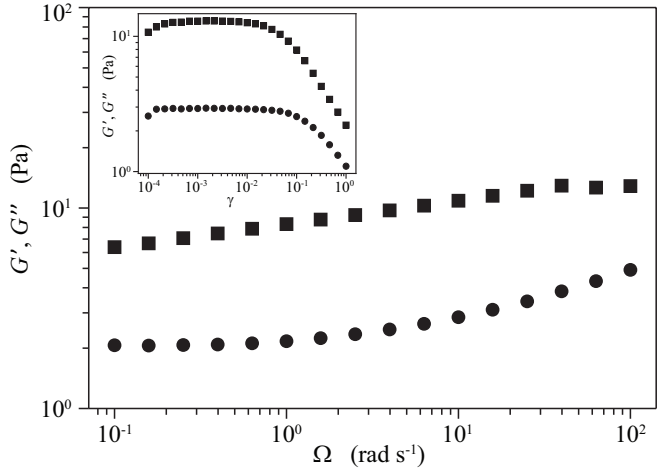}
\caption{Storage modulus $G'$ ($\blacksquare$) and loss modulus $G''$ ($\bullet$) as a function of the frequency $\Omega$ for a constant applied deformation $\gamma=0.02$. Inset: $G'$ and $G''$ as a function of the applied deformation $\gamma$ for a constant frequency $\Omega = 10$~rad~s$^{-1}$.}
\label{Fig04}
\end{figure}

Thereafter, we explore the oscillatory response of a fresh egg white in the same rheological configuration. Amplitude sweeps are performed at a constant frequency $\Omega = 10$~rad~s$^{-1}$, in order to determine the linear viscoelastic region (inset in Fig.~\ref{Fig04}). We choose a deformation within linear viscoelastic regime (\emph{i.e.}, $\gamma=0.02$) to realize the subsequent frequency sweeping and evaluate the elastic/viscous modulus. Figure~\ref{Fig04} shows the storage modulus $G'$ and the loss modulus $G''$ of a fresh egg white as a function of the imposed frequency $\Omega$ for a constant applied deformation $\gamma=0.02$. The mechanical spectrum reveals gel-like properties, as the storage modulus $G'$ is greater than the loss modulus $G''$ over the whole examined range with only slight frequency dependence. These observations are consistent with the response of ovomucin gels at 0.5\% studied by Offengenden and Wu \cite{offengenden2013egg}. Indeed, the glycoprotein ovomucin represents about 3.5\% of the egg white proteins and is characterized by a very high molecular weight of $240 \times 10^6$ at $\rm{pH} \sim 6$; thus, it should strongly contribute to the viscoelastic properties of the egg white.

\begin{figure}[t]
\centering
\includegraphics[width=0.7\hsize]{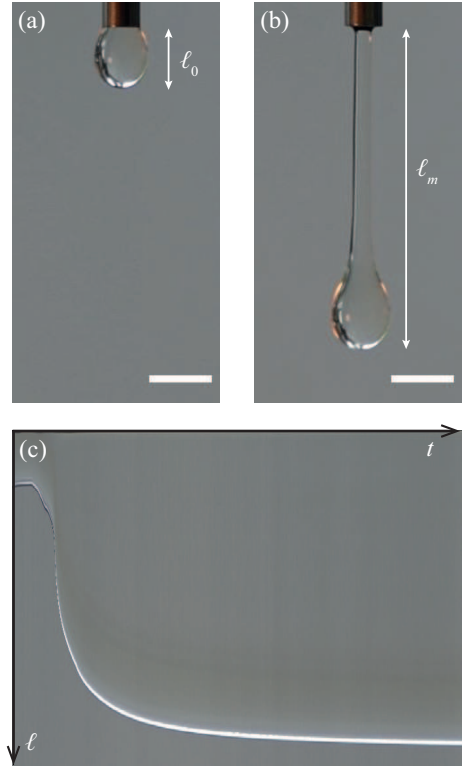}
\caption{Dynamics of an egg white filament: (a)~initial pendant drop of height $\ell_0=2.6$~mm and (b)~final stable filament of hanging length $\ell_m=15.0$~mm. White bars represent 3~mm. (c)~Spatio-temporal diagram of the filament dynamics. The dynamics is observed during 115~s.}
\label{Fig05}
\end{figure}

The gel-like properties of the egg white revealed by rheological measurements are also highlighted outside rheometers by the following simple test: if one places a small volume of egg white between two fingers and pulls them apart, a resulting long filament is observed and remains stable for a few minutes.
The filament exiting a nozzle and hanging a drop is the signature of the molecular weight and conformation of the long molecules as well [Fig.~\ref{Fig01}(b)]. We reproduce this experiment in a more controlled way by applying pressure on the plunger of the syringe to form a pendant drop of egg white as seen in Fig.~\ref{Fig05}(a). While the applied pressure is maintained, the filament extends [Fig.~\ref{Fig05}(b)] and rapidly reaches a stable shape without breaking as evidenced by the spatiotemporal diagram in Fig.~\ref{Fig05}(c).

The large elastic strain withstood by the filament cannot be approached by the theory of linear elasticity. Such deformation can be approached by the neo-Hookean model based on statistical thermodynamics \cite{Treloar}. In general, this model well describes the behavior of elastomers. Hence, the Helmholtz free energy $\mathcal{F}$ represents the work of deformation or elastically stored energy per unit volume of the material. In a three-dimensional description, the work of deformation $W$ expresses as
\begin{equation}
W = C(\mathcal{I}-3),
\label{eq:work1}
\end{equation}
where $C$ is a material constant and $\mathcal{I}$ is the first invariant of the right Cauchy-Green deformation tensor.
$\mathcal{I}$ is expressed as $\mathcal{I}=\lambda_1^2+\lambda_2^2+\lambda_3^2$, where $\lambda_i$ are the stretch ratios along the three main axes and can be larger than 1. Note that the number 3 in Eq.~(\ref{eq:work1}) is included to follow the convention $W = 0$ for $\lambda_i = 1$, with $i=$1, 2 and 3. For constancy of volume, the incompressibility condition writes $\lambda_1 \lambda_2 \lambda_3 = 1$.

In the case of uniaxial loading, \emph{i.e.}, $\lambda_3 = \lambda$, the last condition leads to $\lambda_1 = \lambda_2 = \lambda^{-1/2}$. Hence, the work of deformation per unit volume expresses as
\begin{equation}
W = C\left ( \lambda^2+\frac{2}{\lambda}-3\right ).
\label{eq:work}
\end{equation}
Considering the case of a gravity-induced extension, Eq.~(\ref{eq:work}) allows us to obtain the Helmholtz free energy per unit volume $\mathcal{F}$ \cite{Mora2019}
\begin{equation}
\mathcal{F} \sim \frac{\mu}{2}\left (\lambda^2+\frac{2}{\lambda}-3\right )-\rho g \frac{\ell_0}{2}\lambda,
\label{eq:helmoltz}
\end{equation}
where the inherent elasticity is highlighted by the shear modulus $\mu$. The second term of the right-hand side is the gravity energy density.
The shear modulus is expected to depend significantly on the temperature. At room temperature at which the measurements are performed, the shear modulus is very low because of the weak contribution of the macromolecules stretching.
The following considerations aim to reveal a finite shear modulus for raw egg white in the configuration of the pendant drop.
Noting $\ell_0$ the initial length of the filament and $\ell_m$ the equilibrium length of the filament, the characteristic stretch ratio can be defined as $\lambda = \ell_m/\ell_0$.
The minimization of Eq.~(\ref{eq:helmoltz}) at equilibrium ($\partial \mathcal{F}/\partial\lambda = 0$) gives
\begin{equation}
2\lambda^3-\frac{\rho g \ell_0}{\mu} \lambda^2-2 = 0.
\label{eq:helmoltz_min}
\end{equation}
Applying the predictions of this model to the previous experiment for which we measure $\rho g \ell_0 = 26.5$~Pa and $\lambda= 5.8$, we get an estimate of the shear modulus of the egg white $\mu = 2.3 \pm 0.3$~Pa. This shear modulus is very low compared to usual hydrogels, for which $\mu$ is about 10$^3$~Pa \cite{Subramani}. In this configuration, the obtained value highlights an emergent elasticity. Even if a precise comparison with the storage modulus determined in the rheometer requires to investigate the hanging experiment with further details, the shear and storage modulus are of the same order of magnitude (Fig.~\ref{Fig04}).

\section{Drying cracks}
In the following, we focus on the effect of the long chains constituting the egg white on the drying cracks.
The rheological properties, as the finite value of the shear modulus highlighted above, support the evolution of the mechanical properties during the drying of egg white films. In particular, the emergence of an inherent elasticity during the drying is expected.
At the final stage of the drying, the medium becomes brittle and cracks form. Moreover, the degree of brittleness can be captured through the observation of the cracking [Fig.~\ref{Fig01}(c)].

\subsection{Experimental protocol}
A controlled volume of egg white is deposited at the edge of a glass microscope slide using a micropipette. Then, the film is spread over the entire surface of the slide using a paintbrush.
We take care that the spreading is achieved in one direction, at one time, and at constant velocity $v$, as much as possible. The control of the volume of egg white deposited on the slide and the pressure of application of the paintbrush, although operated manually, enable to obtain films of various thicknesses.
The film thickness is estimated \emph{a posteriori} by differential focus at the film/substrate and film/air interfaces using an optical microscope.
Then, each film naturally dries in the absence of convection in the vapor: the transfer of water in the air is limited by diffusion and is controlled by the relative humidity in the surrounding air.

\subsection{Experimental results}
\begin{figure}
\includegraphics[width=\hsize]{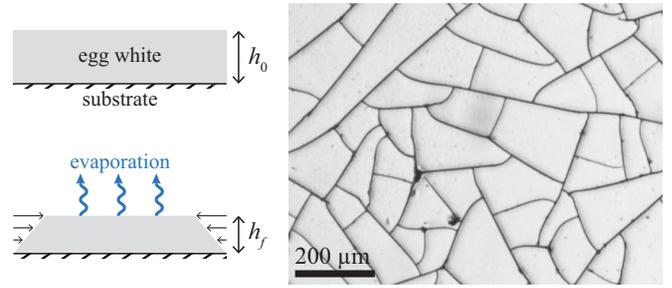}
\caption{Left: side view illustration of the differential shrinkage induced by water removal from an egg white film of initial thickness $h_0$ adhering to a substrate, until complete drying of the film at thickness $h_f$. Right: crack pattern induced by the drying egg white thin film.}
\label{Fig06}
\end{figure}

During the drying process, the formation of a three-dimensional entangled network of long proteins governs the structuration of the egg white film.
Due to the water evaporation, surface capillary forces act on the packing structure that undergoes shrinkage on the drying face.
However, the shrinkage is constrained by the adhesion on the glass slide. This growing misfit results in mechanical tensile stress in the film as sketched in Fig.~\ref{Fig06}\cite{Lazarus2011}. When the stress exceeds a threshold value, cracks invade the film as a result of the stress release.
The cracks are formed successively and finally separate adjacent cells \cite{Bohn2005}.
A typical crack pattern in a film of egg white is shown in Fig.~\ref{Fig06}.

\begin{figure*}
\includegraphics[width=\hsize]{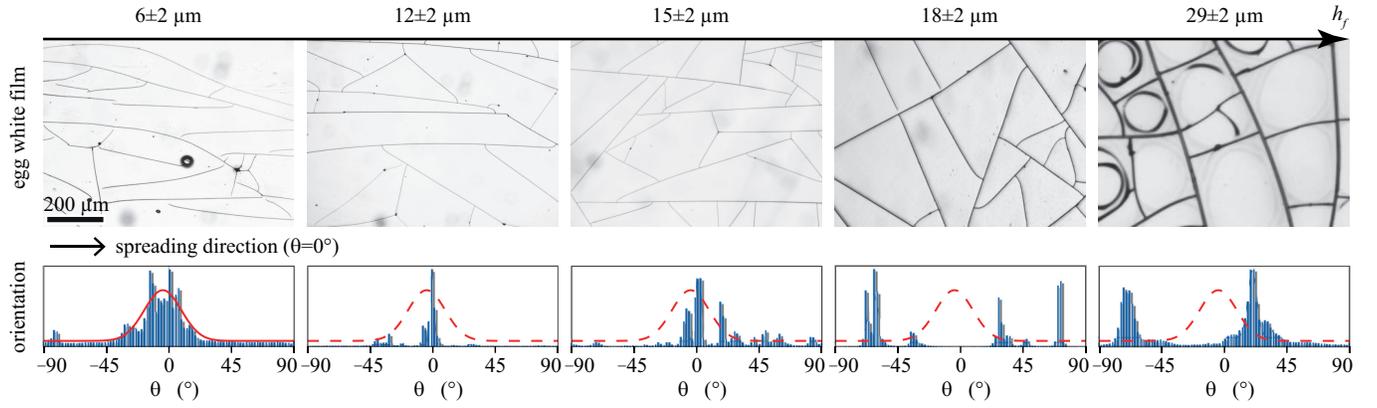}
\caption{Images of crack pattern in films of egg white of different thicknesses. Below each image the statistics on the preferred crack orientation $\theta$ is plotted ($\theta = 0^\circ$ corresponds to the spreading direction). The angle distribution obeys a Gaussian law for thin films (red curve) and then deviates from it as the thickness increases.}
\label{Fig07}
\end{figure*}

Cracks in dried films of egg white are investigated by image processing to infer their connectivity, their width and spacing, and their orientation. This analyze exhibits some of usual characteristics observed in brittle coatings: for example, below a critical thickness $h_c$, the film is crack-free, but for thicker films, junction cracks and then a network of connected cracks preferentially forms \cite{Lazarus2011}. As new cracks keep forming, their length is limited because they grow in a domain limited by the older ones. They collide with them at right angle, forming an increasingly complex polygonal pattern whose properties depend on the film thickness (Fig.~\ref{Fig07}).

However, films of egg white exhibit particular morphologies after the spreading process.
For the two thinner films (above the critical thickness $h_c$), drying cracks exhibit a well-defined orientation, which takes place in the direction of the film spreading ($\theta=0^\circ$), evidenced by a peak in the angle distribution presented in Fig.~\ref{Fig07}. For thicker films, this peak remains predominant, and minor peaks are associated in the distribution. However, various crack orientations are measured for still thicker films overlooking the direction of the film spreading. Finally, the thickest film in Fig.~\ref{Fig07} exhibits a complex morphology made of polygonal cells limited by cracks.
Partial delamination of the cell is made visible by the transparency of the dried film.
The residual mechanical stress in the regions adhering to the substrate can lead to the formation of circular or spiral crack there \cite{Lazarus2011}.

\begin{figure}
\includegraphics[width=0.856\hsize]{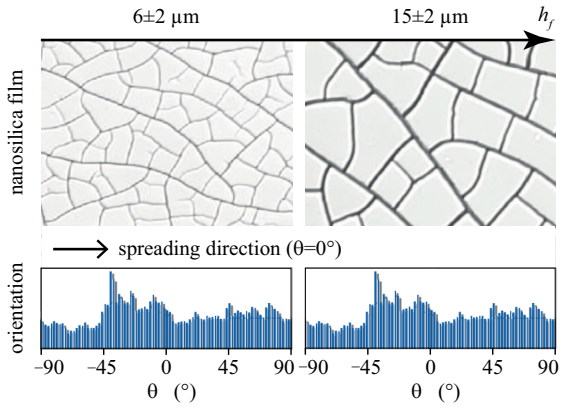}
\caption{Images of crack pattern in films of nearly spherical nanoparticle (Ludox HS40); statistics on the preferred orientation of crack networks is shown for two film thicknesses.}
\label{Fig08}
\end{figure}

The preferred crack orientation exhibited in dried thin films of egg white was explored or in films of particle carrying a magnetic moment under a magnetic external field \cite{Pauchard2008}, or by controlling an external electric field \cite{Tapati2012}, or still in dried films of ellipsoidal nanoparticles \cite{Hisay2018}.
In any event, the formation of cracks reveals the imprint anisotropy of the system.
Nevertheless, the crack pattern induced by the drying of films of spherical nanoparticles does not reveal a preferential direction whatever the film thickness (Fig.~\ref{Fig08}).

\begin{figure}
\includegraphics[width=\hsize]{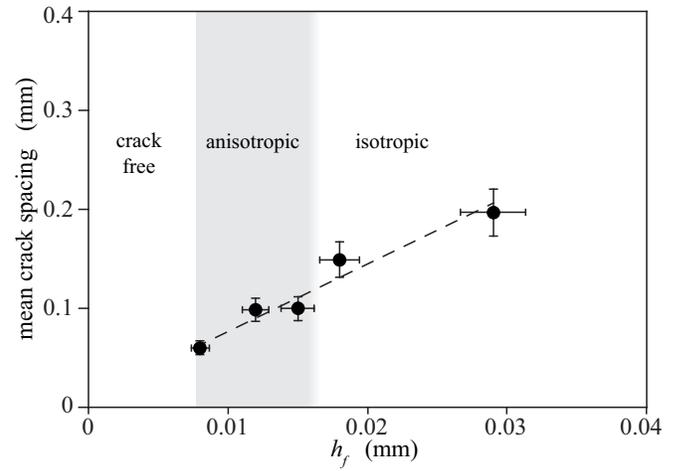}
\caption{Mean crack spacing as a function of the film thickness $h_f$ for egg white. (-~-~-) Linear guideline for the eyes.}
\label{Fig09}
\end{figure}

The mean crack spacing is investigated by image processing and is observed to increase with the thickness of the dried films $h_f$, as shown in Fig.~\ref{Fig09}.
This trend is usually observed for dried particulate films for which the crack spacing increases linearly with the film thickness in the limit of thin films \cite{Lazarus2011,Lazarus2017}.

\subsection{Discussion}
The preferred direction of cracks highlighted in dried thin films of egg white is assumed to be due to the presence of long molecules, imprinting an anisotropy during the solidification.
In general, spreading with the brushstroke causes the layer to shear. In particular, the spreading in one specific direction is a way to align the chains of the egg white along a well-defined direction.
The shear rate resulting from the spreading can be estimated as $\dot{\gamma} = v/h_0$.

\begin{figure}
\includegraphics[width=\hsize]{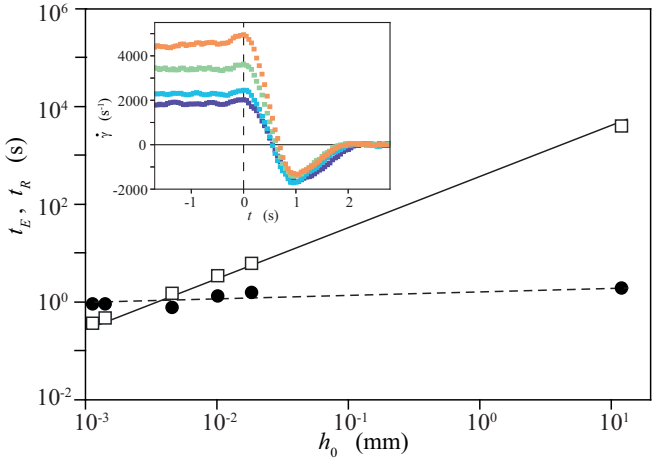}
\caption{Chain relaxation time $t_R$ ($\bullet$) and evaporation time $t_E$ ($\square$) as a function of the initial film thickness $h_0$ in the log-log scale. Inset: Relaxation dynamics of egg white after the shear stress is turned off (indicated by the vertical dashed line at $t=0$) for imposed shear stresses (\textcolor{violet}{$\blacksquare$})~6~Pa, (\textcolor{cyan}{$\blacksquare$})~8~Pa, (\textcolor{green}{$\blacksquare$})~10~Pa, and (\textcolor{orange}{$\blacksquare$})~12~Pa.}
\label{Fig10}
\end{figure}

This allows significant extension simply by chain straightening. However, this alignment can be lost due to chains relaxation since entropy drives the chain recoiling and by the way the recovery of large elastic strains. Hence, two timescales can be considered. First, the relaxation time $t_R$, governing the recoiling process of the chains once the shear has been removed: this timescale depends on the shear rate applied and is extracted from rheological measurements by evaluating its characteristic time of decrease (see inset in Fig.~\ref{Fig10}).
Second the evaporation timescale $t_E$, which controls the film solidification by the drying process: this timescale depends on the film thickness and the evaporation rate $v_E$, as $t_E = h_0/v_E$.
Assuming that the shear process is controlled by the brushstroke moving at the velocity $v \simeq 50$~mm~s$^{-1}$, and the drying process is governed by the evaporation rate $v_E = 2\times 10^{-7}$~m~s$^{-1}$, both timescales are plotted as a function of the film thickness in Fig.~\ref{Fig10}.

If $t_E < t_R$, the alignment of the chains is kept during the film solidification, and the resulting dried film exhibits anisotropic behavior. In this case, the cracks induced by drying exhibit a preferred direction.
However, if $t_E > t_R$, the chains are allowed to lose their alignment by the relaxation process before solidification: the anisotropic behavior of the dried film is not predominant anymore. The resulting cracks induced by drying exhibit an isotropic behavior.
These results highlight the order of magnitude of the film thickness below which the cracks show a well-defined orientation accordingly with Fig.~\ref{Fig07}.

\section{Spin test}
\label{sec:spin}

\subsection{Experimental setup}
\begin{figure}[t]
\includegraphics[width=\hsize]{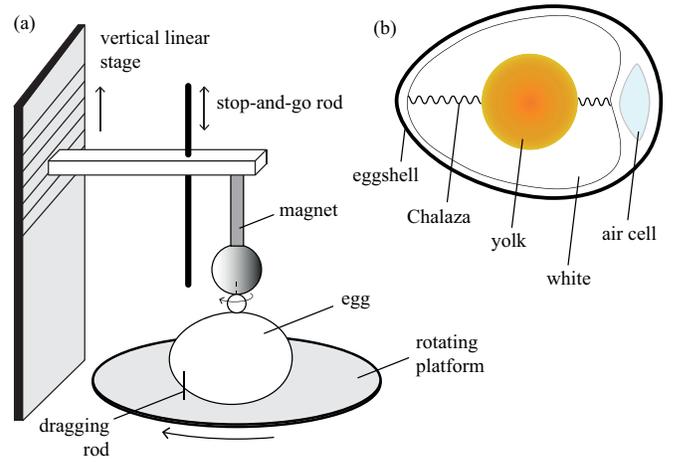}
\caption{(a)~Sketch of the experimental device to reproduce the spin test in the laboratory. (b)~Anatomy of a chicken egg.}
\label{Fig11}
\end{figure}

The experimental setup developed to reproduce the spin test is sketched in Fig.~\ref{Fig11}(a). In order to ensure a rotating system with a low frictional torque, we use two metal spheres of respective diameters 25~mm and 8~mm that are kept in contact by means of a neodymium magnet. The large sphere is kept fixed relative to the magnet, thanks to the strong attractive force of the magnet. The small sphere can rotate almost freely around its contact point with the larger one. A hen egg as sketched in Fig.~\ref{Fig11}(b) (dimension 57 $\times$ 44~mm$^2$ and mass 61~g) is stuck under the small sphere and can be rotated at an angular velocity of 8~rad~s$^{-1}$, thanks to the lower motorized platform. The mechanism for transferring the motion from the platform to the egg is based on a small vertical rod integral with the platform that constraints the egg to follow the platform rotation. Before making any measurement, we ensure that the fluid inside the egg has reached a solid-body rotation, by leaving the egg rotate for several minutes. Then, the magnet is lifted upwards to separate the egg from the lower rotating platform and let it spin freely. The egg is then stopped for a fraction of a second before being released. The stop-and-release procedure is based on a second vertical guided rod that is lowered to intercept the egg trajectory and stops it for a short time prior to be lifted rapidly letting the egg free to rotate again. The ascent and the descent of the rod is done by hand. The dynamics of the egg's rotation is recorded with a video camera with a resolution 1280$\times$960 pixels and frequency 60~Hz. The exact determination of the stopping time is extracted \textit{a posteriori} from video analysis giving a precision of $\pm 1/60$~s. The analysis of the video provides the spinning dynamics of the egg after the stop-and-go procedure. Note that the results presented here were carried out with a single egg, but three other chicken eggs were tested and exhibit completely comparable behavior.

\subsection{Experimental results}

\begin{figure}[t]
\includegraphics[width=\hsize]{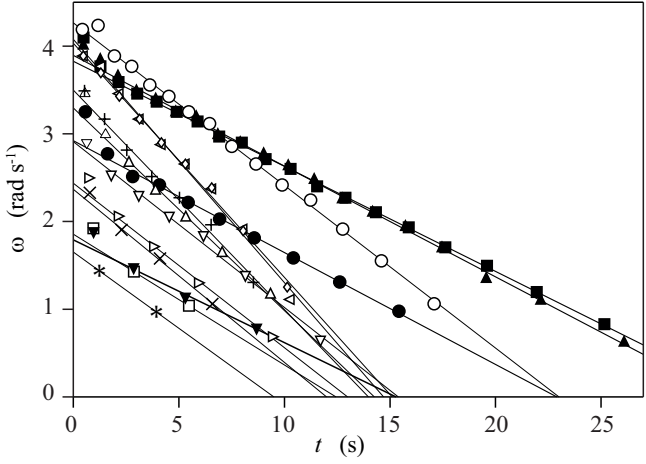}
\caption{Angular velocity $\omega$ of the raw egg as a function of time~$t$, after stopping the egg for a time interval ($\circ$, $\triangleleft$)~$t^*=100$~ms, ($\blacksquare$, $\blacktriangle$)~$t^*=133$~ms, ($\lozenge$)~$t^*=150$~ms, (+)~$t^*=200$~ms, ($\triangle$)~$t^*=217$~ms, ($\bullet$)~$t^*=233$~ms, ($\triangledown$)~$t^*=267$~ms, ($\triangleright$)~$t^*=317$~ms, ($\times$)~$t^*=367$~ms, ($\square$)~$t^*=417$~ms, ($\blacktriangledown$)~$t^*=433$~ms, and ($\ast$)~$t^*=517$~ms. Time $t=0$ corresponds to the release of the egg after having been stopped during $t^*$. Solid symbols correspond to the first four experiments realized and open symbols to the following experiments. Solid lines are the best linear fits of experimental data.}
\label{Fig12}
\end{figure}
When the egg is released, an average angular velocity $\omega$ is calculated each time it makes a half revolution. Figure~\ref{Fig12} displays this angular velocity $\omega$ as a function of time for different stopping times $t^*$. As expected, just after the release at $t=0$, we first observe that the longer the stopping time, the lower the rotational speed of the egg. Thereafter, the angular velocity of the egg is observed to decrease linearly with time, up to the complete stop of the egg. The deceleration of the egg rotation seems to evolve as the number of experiments is increased. Indeed, we note in Fig.~\ref{Fig12} that the angular velocity of the first four experiments performed (solid symbols) decreases more slowly than the following experiments (open symbols). This points out a complex dynamics generated by a possible change in the internal structure of the egg such as the reconfiguration of the tissue band that maintains the yolk (called Chalaza), the displacement of the air cell or the rupture of the membrane surrounding the yolk [Fig.~\ref{Fig11}(b)]. In the following, we will not describe the complete spinning dynamics of the egg, but we rather focus on the rotational speed $\omega_f$ immediately after the release as it shows a good reproducibility over experiments and egg samples. This rotation speed is estimated from linear fits of the data at $t=0$~s.

\begin{figure}[t]
\includegraphics[width=\hsize]{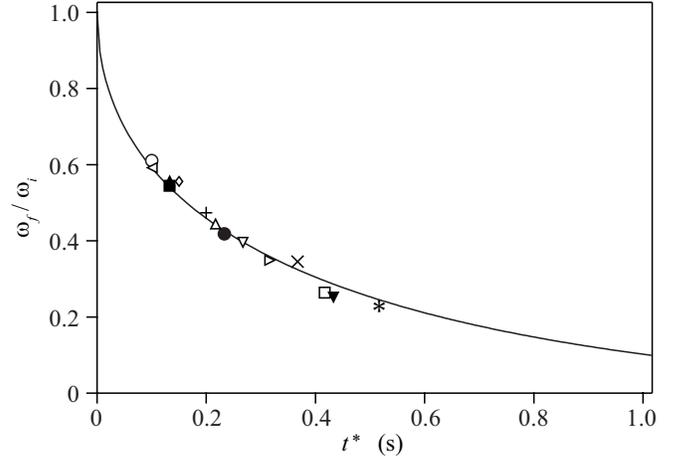}
\caption{Ratio of angular velocities $\omega_f/\omega_i$ after and before stopping the egg rotation as a function of the stopping time $t^*$ (same symbols as in Fig.~\ref{Fig12}). Solid line corresponds to the best fit from Eq.~(\ref{Eq02}), with $R=25$~mm and $\nu=2.2\times 10^{-4}$ m$^2$ s$^{-1}$.}
\label{Fig13}
\end{figure}

The ratio between the releasing angular velocity and the initial angular velocity $\omega_f/\omega_i$ is plotted as a function of the stopping time $t^*$ in Fig.~\ref{Fig13}. This graph shows that $\omega_f/\omega_i$ decreases as the stopping time increases. For a stopping time around 160~ms, the releasing speed of the egg is about half of the initial rotation speed. This behavior is in agreement with our intuition that longer the egg is stopped, larger is the damping of the internal flow, and smaller is the residual rotation after the release.

\subsection{Model}

In order to capture the previous observations, a simple modeling of the problem is proposed. In this approach, the egg can be modeled by a sphere of radius $R$ filled with a homogeneous fluid of density $\rho$. At the start of the experiment, the egg is rotating at an angular velocity $\omega_i$, and the fluid inside is in solid rotation. Under these assumptions, the rotational kinetic energy of the fluid $K=I\omega_i ^2 /2$, where $I$ represents the body's moment of inertia, is equal to $4\pi\rho R^5\omega_i^2/15$. When the rotation is temporarily stopped, the no-slip condition at the wall diffuses toward the center of the fluid and reduces the size of the rotating core. In the limit of low Reynolds numbers, after a stopping time $t^*$, the no-slip condition has invaded a characteristic distance $\delta=\sqrt{\nu t^*}$, where $\nu=\eta /\rho$ is the kinematic viscosity reflecting the momentum diffusivity \cite{guyon2015physical} with $\rho=1037$~kg~m$^{-3}$ for the egg white. The solid rotation of the fluid is now contained in a region of radius $R-\delta$. In the ideal case where the remaining kinetic energy of the rotating core when the egg is released is fully transferred to a new solid rotation of the whole fluid, the new rotation speed $\omega_f$ writes
\begin{equation}
\omega_f= \omega_i \left(1- \frac{\sqrt{\nu t^*}}{R} \right)^{5/2}.
\label{Eq02}
\end{equation}
According to this model, the rotation speed ratio $\omega_f/\omega_i$ is a decreasing function of the stopping time $t^*$, and it becomes null for time longer than $t^*_m = R^2/\nu$. This prediction can be compared to our observations by considering an average radius $R\simeq 25$~mm and letting the viscosity as a free parameter. The best fit of our experiments is found for a dynamic viscosity $\eta \simeq 0.23$~Pa~s, and is presented with a solid line in Fig.~\ref{Fig13}. This estimate allows to evaluate that the thickness of the outer layer affected by the no-slip condition ranges from 19\% to 42\% of the average radius for stopping times between 0.1 and 0.5~s. Considering that the yolk occupies a third of an egg volume (approximated by an ellipsoid), its radius is about 17~mm, which represents 68\% of the average radius $R=25$~mm. These estimates show that for stopping times shorter than 0.5~s, the no-slip condition mainly affects the egg white flow. As a consequence, the lost of rotation speed during a spin test results primarily from the damping of the egg white flow. Furthermore, we note that during the stop of the egg in the spin test, the inner liquid experiences the non-uniform shear rate. However, we can estimate an average value of the shear rate during the time $t^*$, which is $\dot{\gamma}= (R/\sqrt{\nu t^*}-1)\, \omega_i$ . For $R\simeq 25$~mm, $\nu=2.2 \times 10^{-4}$~m$^2$~s$^{-1}$, and stopping times $t^*$ between 0.1 and 0.5~s, the average shear rate $\dot{\gamma}$ ranges from 11 to 35~s$^{-1}$. In this range of shear rates, the rheological measurements provide that the stationary viscosity of the egg white varies between 0.17 and 0.38~Pa~s (see gray area in Fig.~\ref{Fig03}), which is in good agreement with the value estimated from the best fit of spin test experiments with the proposed model, \textit{i.e.}, $\eta \simeq 0.23$~Pa~s.

\section{Conclusion}

In this paper, we realized several experiments with a common chicken egg that can be found in most kitchens. We focus on experiments that involve the liquids contained inside the eggshell and that can be reproduced by everyone: the dynamics of a hanging filament of egg white, the cracks formation in a thin film of egg white after it dries, and the spin test that allows to distinguish between raw and hard-boiled eggs. For each experiment, we point out the main physical ingredients responsible for our observations. We show that the presence of long proteins in the egg white confers a shear modulus to this material and allows the creation of long filaments under stretching. We also evidence a regime of anisotropic crack patterns when a film of egg white is let dried that comes from the alignment of the eggs proteins as the film is deposited. Finally, we show that during a spin test, only the outer liquid layer in the egg is slowed down as a consequence of the viscosity of the egg white, which dissipates part of the rotational energy. The continuation of the flow in the innermost layers explains the residual rotation after the egg release. A simple model allows us to rationalize our observations and provides an estimate of the egg white viscosity that does not disagree with measurements made in a rheometer.

The first approach of these problems proposed in this paper could be the topic of future investigations. In the case of a spinning egg, it would be interesting to rationalize the complete dynamics of the egg after its release and distinguish the influence of several factors such as the presence of the yolk, the surrounding membrane, the chalaza that maintains the yolk, and the presence of air inside the eggshell. More fundamentally, we could study in further detail the rotational dynamics of a sphere filled with a model shear-thinning fluid and propose a complete description for this problem. Also, it would be interesting to complete the rheological characterization of egg white with normal stress measurements in rheometer or through the Weissenberg effect or jet expansion. Finally, the study of the crack appearing in a drying film of egg white would benefit from visualizations of the proteins micro-structure to understand how it impacts the crack pattern.

\begin{acknowledgments}
We are grateful to J.~Amarni, A.~Aubertin, L.~Auffray and R.~Pidoux for their contribution to the development of the experimental setup and C. Manquest for the technical support on rheological measurements. We thank H.~Auradou, A. Davaille and J. Martin for fruitful discussions. B.D.T. was supported by ANR PIA funding: Grant No. ANR-20-IDEES-0002.
\end{acknowledgments}

\section*{Author declarations}
The authors have no conflicts to disclose.

\section*{Data Availability}

The data that support the findings of this study are available from the corresponding author upon reasonable request.

\section*{References}

\bibliography{bertho}

\end{document}